\documentclass[journal,twoside]{IEEEtran}

\usepackage{color}
\usepackage{amssymb}
\usepackage{rotating}
\usepackage[T1]{fontenc}
\usepackage{lscape}
\usepackage{dblfloatfix}
\usepackage{xcolor}
\usepackage{subcaption}
\usepackage[ruled,vlined]{algorithm2e}
\usepackage[autostyle]{csquotes}
\usepackage{tikz}
\usepackage{acro}
\usepackage{siunitx}
\usepackage{enumitem}
\usepackage{multirow}
\usepackage{cite}
\usepackage{hyperref}
\usepackage{amsmath,amssymb,amsfonts}
\usepackage{tikz}
\usepackage{float}
\usepackage{algorithmic}
\usepackage{graphicx}
\usepackage{textcomp}
\usepackage{xcolor}
\usepackage{url}

\usepackage{comment}
\def\BibTeX{{\rm B\kern-.05em{\sc i\kern-.025em b}\kern-.08em
    T\kern-.1667em\lower.7ex\hbox{E}\kern-.125emX}}

\ifCLASSINFOpdf
\else
 
\fi

\usepackage[a4paper, total={180mm,257mm}]{geometry}

\begin{document}

%


\title{Current Topological and Machine Learning Applications for Bias Detection in Text }


%
\author{\IEEEauthorblockN{Colleen Farrelly\IEEEauthorrefmark{1},
Yashbir Singh\IEEEauthorrefmark{2}, 
Quincy A. Hathaway\IEEEauthorrefmark{3}, 
Gunnar Carlsson\IEEEauthorrefmark{4},
Ashok Choudhary\IEEEauthorrefmark{5},
Rahul Paul\IEEEauthorrefmark{6}, 
Gianfranco Doretto\IEEEauthorrefmark{3},
Yassine Himeur\IEEEauthorrefmark{7},
Shadi Atalls\IEEEauthorrefmark{7}, and
Wathiq Mansoor\IEEEauthorrefmark{7}
}\\
\IEEEauthorblockA{\IEEEauthorrefmark{1} Staticlysm LLC, Miami, FL, USA}\\
\IEEEauthorblockA{\IEEEauthorrefmark{2} Radiology, Mayo Clinic, Rochester, MN}\\
\IEEEauthorblockA{\IEEEauthorrefmark{3} West Virginia University, Morgantown, WV, USA}\\
\IEEEauthorblockA{\IEEEauthorrefmark{4}Stanford University, California, USA}\\
\IEEEauthorblockA{\IEEEauthorrefmark{5}Department of surgery, Mayo Clinic, Rochester, MN, USA}
\\IEEEauthorblockA{\IEEEauthorrefmark{6}Harvard University, USA}
\\IEEEauthorblockA{\IEEEauthorrefmark{7}College of Engineering and Information Technology, University of Dubai, Dubai, UAE}
}


\maketitle

\begin{abstract}
Institutional bias can impact patient outcomes, educational attainment, and legal system navigation. Written records often reflect bias, and once bias is identified; it is possible to refer individuals for training to reduce bias. Many machine learning tools exist to explore text data and create predictive models that can search written records to identify real-time bias. However, few previous studies investigate large language model embeddings and geometric models of biased text data to understand geometry’s impact on bias modeling accuracy. 
To overcome this issue, this study utilizes the RedditBias database to analyze textual biases. Four transformer models, including BERT and RoBERTa variants, were explored. Post-embedding, t-SNE allowed two-dimensional visualization of data. KNN classifiers differentiated bias types, with lower k-values proving more effective. Findings suggest BERT, particularly mini BERT, excels in bias classification, while multilingual models lag. The recommendation emphasizes refining monolingual models and exploring domain-specific biases.
\end{abstract}


\IEEEpeerreviewmaketitle

\section{Introduction}
In recent years, the intersection of topological data analysis and machine learning has opened up exciting new avenues for understanding and addressing the issue of bias in text data \cite{himeur2021survey,sardianos2021emergence,singh2023role}. With the proliferation of textual information on the internet, the potential for bias, both subtle and overt, has become a pressing concern in natural language processing (NLP) and text analysis \cite{farhat2023analyzing,singh2023topological}. Bias can manifest in various forms, such as gender bias, racial bias, or political bias, and can significantly impact the fairness and accuracy of text-based systems, including search engines, recommendation systems, and sentiment analysis tools \cite{himeur2022latest,sardianos2020rehab,borkan2019nuanced}.

\subsection{Stigma and Bias}
Stigma is a perceived identity within a society or subgroup arising from a mismatch between social identities valued by those in power within a society and a person’s true identity \cite{1}. Examples include physical characteristics (such as race or missing limb differences), medical disorders (such as substance use disorders or visual impairment), and changeable visual cues of subgroups (such as tattoos or mohawks within a society where the majority and those in power do not display these visual cues). Stigma can be associated with two forms of bias: 1) implicit biases, where the person within a majority group acts with bias but is unaware that they are acting on their bias against the stigmatized group \cite{2}, and 2) explicit biases, where the person within the majority group acts with bias and is aware that they are acting on their bias against the stigmatized group \cite{3}. Thus, stigmatized groups often face unique societal challengesbecause of implicit and explicit biases, whichcan happen formally or informally.

\subsection{Institutional Bias}
Characteristics devalued by those in power or by the majority in  society as different than the society’s norms become codified into laws and regulations within societal institutions, and individuals with those characteristics often face institutional barriers \cite{1}. For instance, a student with visual impairment in an educational system that assumes all students can see will run into problems with many academic tasks, like reading assignments or the blackboard, unless another option is created; fortunately, this is usually the result of an implicit bias that is quickly identified by the educational system \cite{4}. Jim Crow laws in the American South exemplify explicit bias within the legal system \cite{5}.

\subsection{Non-Institutional Bias}
Devaluation of individuals based on characteristics can also be more informal with respect to everyday interactions. For instance, a physician treating a patient with a history of substance use disorder which presents with pain may quickly label the patient as “drug-seeking,” which then follows the patient throughout their journey to receive a diagnosis \cite{6,7}; hopefully, this is an implicit bias and not an intentional dismissal of a patient. Implicit bias against female patients has been well-documented as leading to adverse outcomes for patients presenting in emergency settings with symptoms that diverge from androcentric cardiac symptoms in emergency settings \cite{8,9,10}. Both types of informal biases often become codified in the clinical record by physicians, creating institutional barriers through documented language by those in authority \cite{11,12,13,14}

\subsection{Implicit Bias Solutions}
Because implicit bias is not intentional, antibias training work effectively in combatting bias and have lasting effects \cite{15,16}. Within the medical setting, implicit biases against patients can be ameliorated through diversity training, such as in Morris et al., 2019, which provided a curriculum for working with LGBTQIA+ patients for current students. Interventions like this can likely assuage the issues in other fields, such as the legal field or education. Creating inclusive environments has also been shown to combat biases \cite{17}.

\subsection{Natural Language Processing}
One branch of machine learning, natural language processing (NLP), processes and analyzes text data \cite{sohail2023future}. Many tools exist in this field, but within the context of bias detection, a few tools are more relevant. Typically, a block of text is first parsed into individual words via a process called tokenization \cite{sohail2023using,sohail2023decoding}. Then, each token can be matched to terms of interest, through a pre-trained algorithm that recognizes terms of interest (such as famous people or places) or a custom matching dictionary. However, for large document sets with an extensive list of terms for which to search, this can be computationally intensive \cite{farhat2023analyzing,khennouche2024revolutionizing}.
Another tool commonly used to wrangle text data into matrix form for use in supervised learning models, such as logistic regression or random forest classifiers, is embedding text. Many ways to do this exist. First, one can simply encode the full sets of documents by frequency of each term that exists at least once in the set of documents, usually according to preset transformations or weightings of frequencies across words and across documents \cite{18}; term frequency–inverse document frequency (TF-IDF) is one common method to transform text documents \cite{18}. Many other embeddings exist, including word2vec and GloVe \cite{19,20}. One drawback of this method is the potential dimensionality of the matrix, which can become problematic for statistical models or machine learning algorithms modeling the data from the embedding matrix.
Another embedding option is to use a pre-trained model that can embed text according to the text meaning and map these meanings to a matrix with a pre-defined size to limit the potential size of the matrix after embedding all the text documents. Bidirectional Encoder Representations from Transformers (BERT) is one such option; implementations of BERT typically use an algorithm to map new text to the pre-trained embedding \cite{21}. This process can be computationally intensive, though, and it can require additional computing resources to obtain the full document set embedding; further, matrices typically don’t have as many potential predictors as a TF-IDF embedding.
BERT has already been used to detect online hate speech \cite{22,23}. In addition, some studies have shown racial and gender bias in word embeddings themselves, so such a method may not be suited for bias detection applications \cite{24}. In addition, hundreds of BERT variations exist, and it is not known how different variations perform with respect to bias detection.

\subsection{Machine Learning on Text Embeddings}
Once text data is embedded through TF-IDF, BERT, or other methods, machine learning, and statistical models can be used to explore and predict biases within the text documents. Supervised learning, where the machine learning model learns to predict a particular outcome (such as spam/not spam or biased/not biased), is a common task in text analytics \cite{25,kheddar2023deep}. Models are trained on a set of embedded text documents, followed by validation on the remaining text documents to ensure the model can generalize to other documents. Many studies have examined ways to classify racial or gender bias within document sets \cite{22,23,26}.
k-nearest neighbor (KNN) classifiers are common in text classification problems and rely on local geometry to classify a point based on the points nearest that point in the geometry of the embedding space \cite{27,28,29,30}. For the classification of technical documents, Larkey and Croft (1996) found advantages of KNN models and embedding strategies; given the linguistic nuances of biased language, it is possible that embedding strategies coupled with KNN models will work well for bias detection.
Unsupervised learning, which includes data mining methods like clustering or visualization, can also be useful in text bias detection, particularly when specific terms are not known ahead of time. Unsupervised learning on text data has been used to understand what exists in qualitative data \cite{31}, to explore topics in text datasets, and to summarize texts \cite{32}, among many other applications. Unsupervised learning may be a good first step in bias detection pipelines, as it can uncover bias types in clinical notes.
T-stochastic neighbor embedding (t-SNE), a dimensionality reduction technique that creates pairwise probability distributions to embed points in lower-dimensional spaces, has been used in text visualization \cite{33,34}. Given t-SNE’s ability to visualize high-dimensional text embeddings in low-dimensional space, it is likely that t-SNE will make a good tool for exploring bias separation in text embeddings such as BERT.

\section{Methods}
Data consisted of text samples curated in the RedditBias database (\url{https://github.com/SoumyaBarikeri/RedditBias}), containing religious, racial, gender, and orientation bias \cite{35}. Because religious bias had a larger text sample (2139 for religious. 504 for each of the others) than other bias types, we employed random sampling on the religious bias dataset to create a similar-sized subsample of religious bias (504 examples) compared to other bias subsamples (504 examples each in racial, gender, and orientation bias).

Four types of pre-trained transformer models were fit on the RedditBias samples. Our first BERT model was the all-mpnet-base-v2 (full BERT) model from the HuggingFace repository of Python’s sentence\_transformer package \cite{36}, which is a sentence embedding version of BERT trained on 1 billion sentence pairs. Our second BERT model was all-MiniLM-L6-v2 model (mini BERT), which uses the same training set as our full BERT model but embeds the data into a smaller space (creating a dense embedding). Our first RoBERTa model was the all-roberta-large-v1 model (all RoBERTa), which uses the same training data with the RoBERTta design rather than BERT design of transformer and a sparser, high-dimensional embedding space. Our second RoBERTa model was the xlm-roberta-base model (raw RoBERTa), which trains on 2.5 terabytes of scraped data across 100 languages and embeds result in a smaller space with a denser embedding without any data curation or validation prior to training the model.

\begin{figure*}[htbp]
\centering
\includegraphics [width=0.78\textwidth]{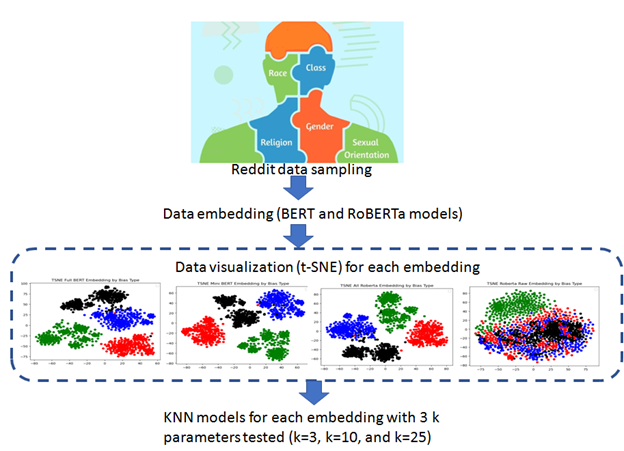}
\caption{Flow of data processing from sampling to embedding with BERT models to visualizing embeddings to creating KNN models with varying k parameter.}
\label{fig1}
\end{figure*}

After embedding the data, t-SNE was applied to the embeddings to reduce dimensionality to two dimensions for easy visualization. As mentioned, t-SNE is a dimensionality reduction method that relies on pairwise probability distribution calculation to embed pairs of points into a lower-dimensional space \cite{37} and falls into the broad field of dimensionality reduction techniques. We implemented t-SNE (Default parameters of TSNE from sklearn manifold package with n\_components=2) on our four embeddings through scikit-learn’s Python implementation of t-SNE \cite{38}.
KNN classifiers classify point labels based on the k points nearest a given point through label voting \cite{39}. If k=5 and four points are label X while one point is label Y, the point in question would be assigned the label X, as more votes from neighbors were for the label X. Thus, local geometry and point distributions play a large role in the performance of KNN models. Distance metric choice can play a large role in performance \cite{40}. The choice of k can lead to either underfitting (larger k values) or overfitting (smaller k values) of a model, with k acting as a smoothing parameter \cite{41}. We fit our KNN classifiers with sci-kit-learn’s Python implementation of t-SNE, a varying choice of k (3, 10, or 25), a Euclidean distance metric (which is native to BERT and RoBERTa embeddings), and the four embeddings themselves as algorithm input (rather than their reduced forms in the t-SNE application) (Fig. \ref{fig1}).
KNN classifiers were run 50 times each to compare results. Bonferroni-corrected t-tests then served to distinguish classifier performance statistically across runs.

\section{Results}
\begin{figure}[htbp]
\includegraphics [width=0.38\textwidth]{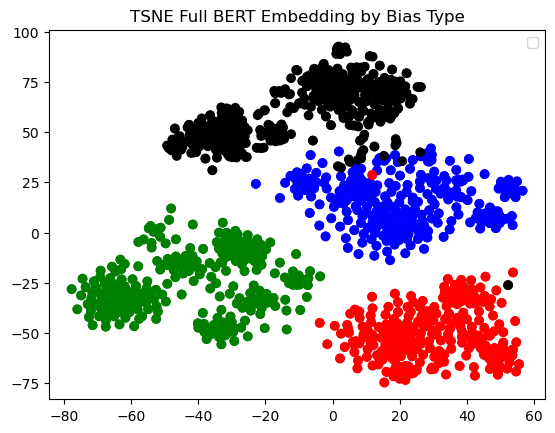}
\caption{t-SNE embedding of all-mpnet-base-v2 model of our Reddit sample.}
\label{fig2}
\end{figure}

\begin{figure}[htbp]
\includegraphics [width=0.38\textwidth]{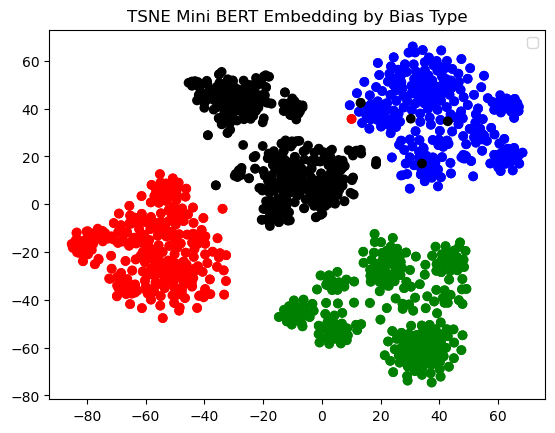}
\caption{t-SNE embedding of all-MiniLM-L6-v2 model of our Reddit sample.}
\label{fig3}
\end{figure}

\begin{figure}[htbp]
\includegraphics [width=0.38\textwidth]{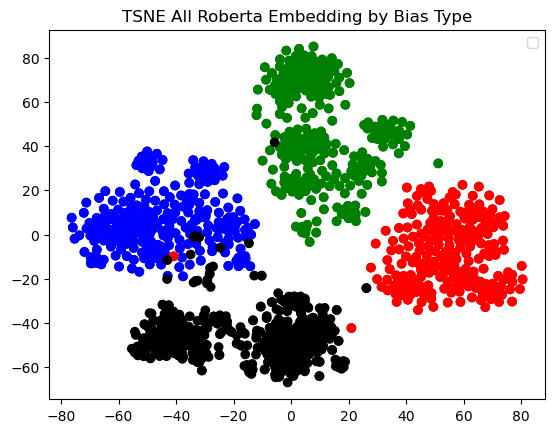}
\caption{t-SNE embedding of all-roberta-large-v1 model of our Reddit sample.}
\label{fig4}
\end{figure}

\begin{figure}[htbp]
\includegraphics [width=0.38\textwidth]{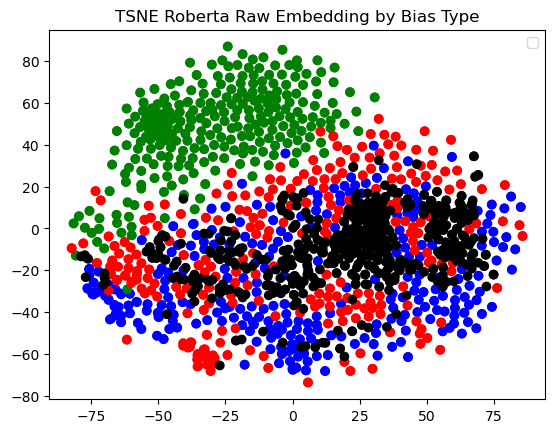}
\caption{t-SNE embedding of xlm-roberta-base model of our Reddit sample.}
\label{fig5}
\end{figure}

The t-SNE plots suggest that most of our chosen embeddings separate bias types well (Fig. \ref{fig2}, Fig. \ref{fig3}, Fig. \ref{fig4}, and Fig. \ref{fig5}). The raw text RoBERTa model that included many language examples did not perform as well as the other three embeddings, suggesting that a manual review of training data and focus on embeddings that only include the language of interest may be useful in the context of bias detection. BERT embeddings may perform better than RoBERTa embeddings in this context, as well. For the best embeddings, such as our full BERT embedding, almost no overlap exists, suggesting that KNN models would perform very well on the classification task.

Models with k=3 and k=10 performed better across embeddings (70\% training, 30\% test)than k=25 (p<0.01), and BERT embeddings showed superior performance to RoBERTa embeddings across choices of k (p<0.01), with the mini BERT embedding showing the best overall performance, particularly at k=3 (p<0.01) (Table \ref{tab1}).

Table \ref{tab1}. Embedding type and number of nearest neighbors impacted KNN classifier performance.
 Our sample data included misspellings, grammatical mistakes, and varying text lengths. These results suggest that our chosen BERT models and the single-language RoBERTa model are robust to real-world English-language text, including text with misspellings and grammatical errors. However, multilingual models without validation or cleaning of training data do not seem to work as well, though they are currently the only option for multilingual document sets.

\begin{table}[t!]
\centering
\caption{Embedding Results}
\label{tab1}
\begin{tabular}{l|c|c|c}
\hline
Embedding & \(K=3\) & \(K=10\) & \(K=25\) \\
\hline
Full BERT & \(0.99 (0.98,1.00)\) & \(0.99 (0.97,1.00)\) & \(0.99 (0.97,1.00)\) \\
Mini BERT & \(1.00 (0.99,1.00)\) & \(0.99 (0.98,1.00)\) & \(0.99 (0.98,1.00)\) \\
Full RoBERTa & \(0.99 (0.98,1.00)\) & \(0.99 (0.98,1.00)\) & \(0.99 (0.97,0.99)\) \\
Raw RoBERTa & \(0.87 (0.84,0.90)\) & \(0.88 (0.82,0.91)\) & \(0.88 (0.83,0.91)\) \\
\hline
\end{tabular}
\end{table}

\section{Discussion}
Logical next steps include pilot studies of applying this methodology to educational notes, medical notes, and legal notes to assess the accuracy and feasibility of detecting bias in these sources based on the proposed embedding strategy coupled with KNN models. It is possible that embeddings will not work as well in fields with substantial jargon, such as the medical field, and testing field-specific embeddings may be worth trying. Further, the medical field, including the electronic medical record and clinical notes, likely has lower frequencies of overt racial, religious, gender, and orientation bias. Applying our proposed model within healthcare likely would include tailoring the algorithm to identify more difficult forms of bias, such as negative patient descriptors [11]; these could include terms such as aggressive, agitated, angry, challenging, combative, etc.

In addition, the assessment of value-added and the potential human impact of applying these classifiers to real-world institutional data should be a part of these pilot studies. However, given the promising results of this study, it is worth exploring this methodology on field-specific text data. Reliable detection of different types of bias can pinpoint areas of improvement for institutions and individuals and pinpoint what type of training would be needed to reduce bias. Reduced institutional bias can, in turn, improve outcomes for those interacting with institutions.

Because our data involves only Reddit (Reddit Inc © 2023) samples, it is hard to know how the results would generalize to longer text documents or to text documents with field-specific jargon. Because prior studies have shown that good embeddings plus a KNN model can improve text classifier accuracy [29], it is likely that the methodology will generalize, if not the specific embeddings that worked well in this study. Further research is needed to determine ideal embeddings that capture the jargon and biased language within specific contexts, such as clinical notes or legal briefs.

For now, we suggest using monolingual embedding models with some validation and cleaning of input data for embedders, as multilingual models without any curation/validation steps did not give good results. It is possible that better multilingual models will exist in the future, and multilingual and multicultural bias applications will be more feasible at that point in time.

\section{Conclusion}
In our study, we aimed to understand the efficacy of pre-trained transformer models in detecting various biases in textual data, specifically from the RedditBias database. Our findings indicate that BERT embeddings, particularly the mini BERT embedding, outperformed RoBERTa embeddings in classifying bias types. The performance was especially notable with certain choices of the '$k$' parameter in KNN models, with $k$=3 yielding the best results. It was observed that the raw RoBERTa model, trained on a large multilingual dataset, was less effective compared to other embeddings, emphasizing the potential importance of data curation and language specificity in training models for bias detection.

The results highlight the robustness of the BERT models and the single-language RoBERTa model in handling real-world English text nuances, such as misspellings and grammatical errors. However, multilingual models that lack a validation or cleaning phase in their training data exhibited reduced effectiveness. This suggests that while there's potential for future multilingual models to improve, the current recommendation would be to employ monolingual embedding models with validated and curated input data.

Given the success of the models on Reddit data, there's optimism about applying the methodology to field-specific texts, such as educational, medical, and legal notes. However, careful consideration is required due to potential challenges posed by field-specific jargon and subtler forms of bias. Regardless, the methodology offers the potential for institutions to identify and rectify biases, ultimately improving outcomes for the diverse communities they serve.

Looking forward, as technologies and embeddings evolve, future research should explore optimized embeddings to capture nuanced and context-specific biased language.

\section{Compliance with Ethical Standards}
Ethical and informed consent for data used: Not Applicable
\section{Funding}
No funding
\section{Competing Interests}
The authors declare that they have no conflict of interest.
Data availability and access: The datasets created and/or analyzed during the current investigation are available upon reasonable request from the corresponding author.

\section*{Acknowledgment}


\end{document}